\documentclass[9pt,a4paper,twocolumn]{article}

\usepackage[utf8]{inputenc}
\usepackage{dcolumn}
\usepackage{bm}
\usepackage{braket}
\usepackage[T1]{fontenc} 
\usepackage{float}
\usepackage{amsmath,amsfonts,amssymb,eqnarray}  
\usepackage{color,soul}
\usepackage{graphicx}  
\usepackage{array,booktabs}
\usepackage[pdftex,hypertexnames=false]{hyperref} 
\usepackage[affil-it]{authblk}
\usepackage[font=small,labelfont=bf]{caption}
\usepackage{cite}
\usepackage[left=2cm,right=2cm,top=1.6cm]{geometry}
\usepackage{titlesec}
\usepackage{multirow}
\definecolor{mycol}{RGB}{255,230,204}
\usepackage{mdframed}
\usepackage[binary-units = true]{siunitx}
\usepackage{multirow, makecell}
\newcommand{\red}{\color{black}}
\usepackage[normalem]{ulem}
\titleformat{\section}{\normalfont\fontsize{13}{15}\bfseries}{\thesection}{0.5em}{}

\title{\textbf{Towards fully-fledged quantum and classical communication over deployed fiber with up-conversion module}
} 

\author{Davide Bacco$^{1* \dagger}$, Ilaria Vagniluca $^{2,3 \dagger}$, Daniele Cozzolino$^{1}$, Søren M. M. Friis$^{4}$, Lasse Høgstedt$^{4}$, Andrea Giudice$^{5}$, Davide Calonico$^{6}$, Francesco Saverio Cataliotti$^{3,7,8}$,\\ Karsten Rottwitt$^{1}$, Alessandro Zavatta$^{3,7,8}$}

\affil{\small{
$^\textrm{\textit{1}}$Center for Silicon Photonics for Optical Communication (SPOC), Department of Photonics Engineering, Technical University of Denmark, 2800 Kgs. Lyngby, Denmark.\\
$^\textrm{\textit{2}}$ Department of Physics “Ettore Pancini", University of Naples “Federico II", 80126 Naples, IT\\
$^\textrm{\textit{3}}$CNR - Istituto Nazionale di Ottica (CNR-INO), Largo E. Fermi, 6 - 50125 Firenze, Italy.\\

$^\textrm{\textit{4}}$NLIR ApS, Hirsemarken 1 1st floor, 3520 Farum, Denmark.\\
$^\textrm{\textit{5}}$Micro Photon Devices S.r.l., via Antonio Stradivari 4, 39100 Bolzano, Italy\\
$^\textrm{\textit{6}}$ I.N.Ri.M. Istituto Nazionale di Ricerca Metrologica, Torino,Italy\\
$^\textrm{\textit{7}}$LENS and Dipartimento di Fisica e Astronomia, Università di Firenze, Via G. Sansone, 1 - 50019 Sesto Fiorentino, Italy.\\
$^\textrm{\textit{8}}$ QTI SRL, Largo Enrico Fermi, 6 - 50125 Firenze, Italy.\\
$\dagger$ These authors contributed equally to this work\\
* dabac@fotonik.dtu.dk
}}
\date{}

\begin{document}

\maketitle

\begin{figure*}[ht!]
\centering
\includegraphics[width=0.99\textwidth]{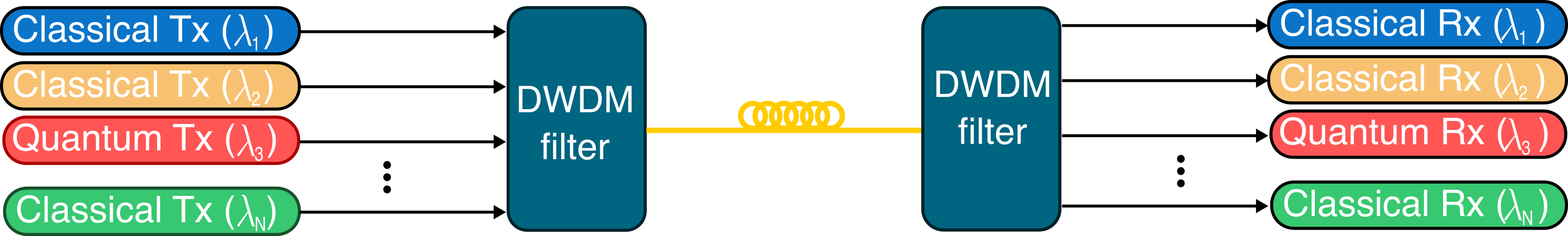}
\caption{\textbf{Schematic of classical and quantum communication {\red channels} in a dense wavelength multiplexing scheme.} Multiple classical transmitters are combined with a quantum one in a dense wavelength division multiplexing scheme. After the propagation through the communication channel{\red{,}} the different wavelengths are separated and measured.}
\label{fig:idea}
\end{figure*}

\begin{abstract}
Quantum key distribution (QKD), the distribution of quantum secured keys useful for data encryption, is expected to have a crucial impact in the next decades. 
However, although the notable achievements accomplished in the last twenty years, many practical and serious challenges are limiting the full deployment of this novel quantum technology in the current telecommunication infrastructures. In particular, the co-propagation of quantum signals and high-speed data traffic within the same optical fiber, is not completely resolved, due to the intrinsic noise caused by the high-intensity of the classical signals.  
{\red{As a consequence,}} current co-propagation schemes limit the amount of classical optical power in order to reduce the overall link noise. {\red{However,}} this ad-hoc solution restrains the overall range of possibilities for a large scale QKD deployment.
Here, we propose and demonstrate a new method, based on up-conversion assisted receiver, for co-propagating classical light and QKD signals. In addition, we compare the performances of this scheme with an off-the-shelf quantum receiver, equipped with a standard InGaAs detector, over different lengths of an installed fiber link. Our proposal exhibits higher tolerance for noise in comparison to the standard receiver, thus enabling the distribution of secret keys in the condition of 4 dB-higher classical power.
\end{abstract}
\section*{Introduction}
Quantum key distribution (QKD) aims at distributing unconditional secure keys, useful for protecting our data communications~\cite{pirandola2020advances,xu2020secure}. By exploiting the properties of quantum mechanics, it is possible to deliver information theoretic secure keys which can be used for the encryption and decryption of data and messages.
A conspicuous number of implementations (e.g., discrete variable, continuous variable, differential phase-shift modulation) and in-field experiments have been carried out in the last decades, demonstrating the feasibility of such technology~\cite{shimizu2014,tang2016,collins2016,zhang2017,bacco2019boosting,avesani2020,bacco2019fi}. However, multiples open problems are still limiting the full deployment of QKD in real-world applications. For example, the low key generation rate and the co-existence with the already-existing infrastructure for optical communication, are the central points for a successful implementation of this technology on a large-scale. The first limitation can be surmount{\red{ed}} either by improving the state-of-art devices (i.e., by employing higher repetition rate transmitters or better performing single-photon detectors), or by adopting novel quantum communication schemes with multidimensional modulation~\cite{cozzolino2019high}. In particular, multiple degrees of freedom of light can be employed simultaneously to enlarge the Hilbert space dimensions, thus increasing the information capacity of single photons and enhancing the secret key rate~\cite{mower2013high,bunandar2015practical,Steinlechner2017,Martin2017,Wang2018,Ding2017}. Among the various degrees of freedom to be adopted for quantum communication, time-bin encoding is the most suitable for propagation in single-mode fiber links~\cite{islam2017provably,vagniluca2020}. 
The second open problem, the compatibility with existing telecommunication infrastructures, is still very challenging although many solutions have been tested. 
For example, in order to co-propagate classical and quantum signals within the same fiber, various approaches can be adopted: time-division multiplexing, space-multiplexing, polarization multiplexing and wavelength division multiplexing (WDM)~\cite{tanaka2008,choi2014,mao2018,gustavo2020}.
Both time and space approaches are interesting and promising for future development, but the most common and versatile in the current telecommunication networks is the frequency multiplexing for serving multiple users~\cite{peters2009dense}. 
In the WDM technique, different wavelengths in the C-band are used for sending different data or to distribute communication between different users. The general idea is depicted in Figure~\ref{fig:idea}, in which multiple transmitters (Classical TX $\lambda_{1, ..., N }$) are combined in a dense wavelength division multiplexer filter and sent over an optical fiber link. Within the $N$ transmitters, $\lambda_{3}$ is used for quantum communication and in particular for QKD. At the other end of the fiber link, a similar filter is used to separate all the {\red{different}}  wavelengths. Although this configuration is very convenient from a practical point of view, the quantum signal {\red seriously} suffers from the proximity of the high-intensity classical light, which generates a large amount of extra noise and thus it limits the overall performance of the quantum communication. 
To be more precise, the interaction of high-intensity laser with the optical devices and fiber link  generates photons at different wavelength{\red{s}} (scattering Raman, Brillouin and Rayleigh) which can survive to the DWDM filter{\red{, thus resulting in}} a source of noise for the fragile quantum signal ~\cite{eraerds2010}. 
In addition, the standard components for optical communication (i.e., optical filters, attenuators, isolators, etc..) are designed and tested for classical intense signals, which have less requirements in terms of loss and extinction ratio. We believe that a proper design of new components (quantum custom components) would help in the development of large-scale quantum networks. 
As a matter of that{\red{,}} the co-propagation of classical and quantum light is very challenging and the most common and practical solution is lowering the amount of classical power which is injected in the fiber, in order to reduce the scattering effects.
However, this solution cannot always be adapted since in many classical networks the required input power is 0 dBm, which is the standard value for classical optical communication. 
{\red{An alternative}} solution is represented by the wide-range wavelength multiplexing, in which is possible to combine the already existing data traffic in the C-band with a quantum transmitter in the O-band, as demonstrated in previous works~\cite{mao2018}. {\red Another effective approach is to employ continuous-variable (CV) QKD protocols, where homodyne and heterodyne detection schemes are exploited instead of single-photon detectors, thus enabling a powerful and intrinsic filter of the optical mode related to quantum signals~\cite{qi2010feasibility,kumar2015coexistence,eriksson2019wavelength}. However, CV-QKD is still limited in terms of secret key rate achievable and transmission distances, as well as of practical security proofs, with respect to the more advanced and long-studied discrete-variable (DV) QKD. Therefore,} the challenge of exploiting the existing infrastructure for dense wavelength division multiplexing of quantum and classical signals in the C-band, {\red for DV-QKD protocols,} is still unsolved. 
In this work, we propose and demonstrate the possibility of co-propagating classical and quantum light through adjacent channel{\red{s}} in a dense-wavelength multiplexing approach, by exploiting the intrinsic filter of a frequency up-conversion {\red receiver} at the single-photon level, combined with the timing performances of silicon single-photon avalanche detector. 
In addition, we compared our new scheme with an off-the-shelf quantum receiver, equipped with a standard indium gallium arsenide (InGaAs) detector, over different channel lengths of a metropolitan deployed fiber . 

\begin{figure*}[ht!]
\centering
\includegraphics[width=0.99\textwidth]{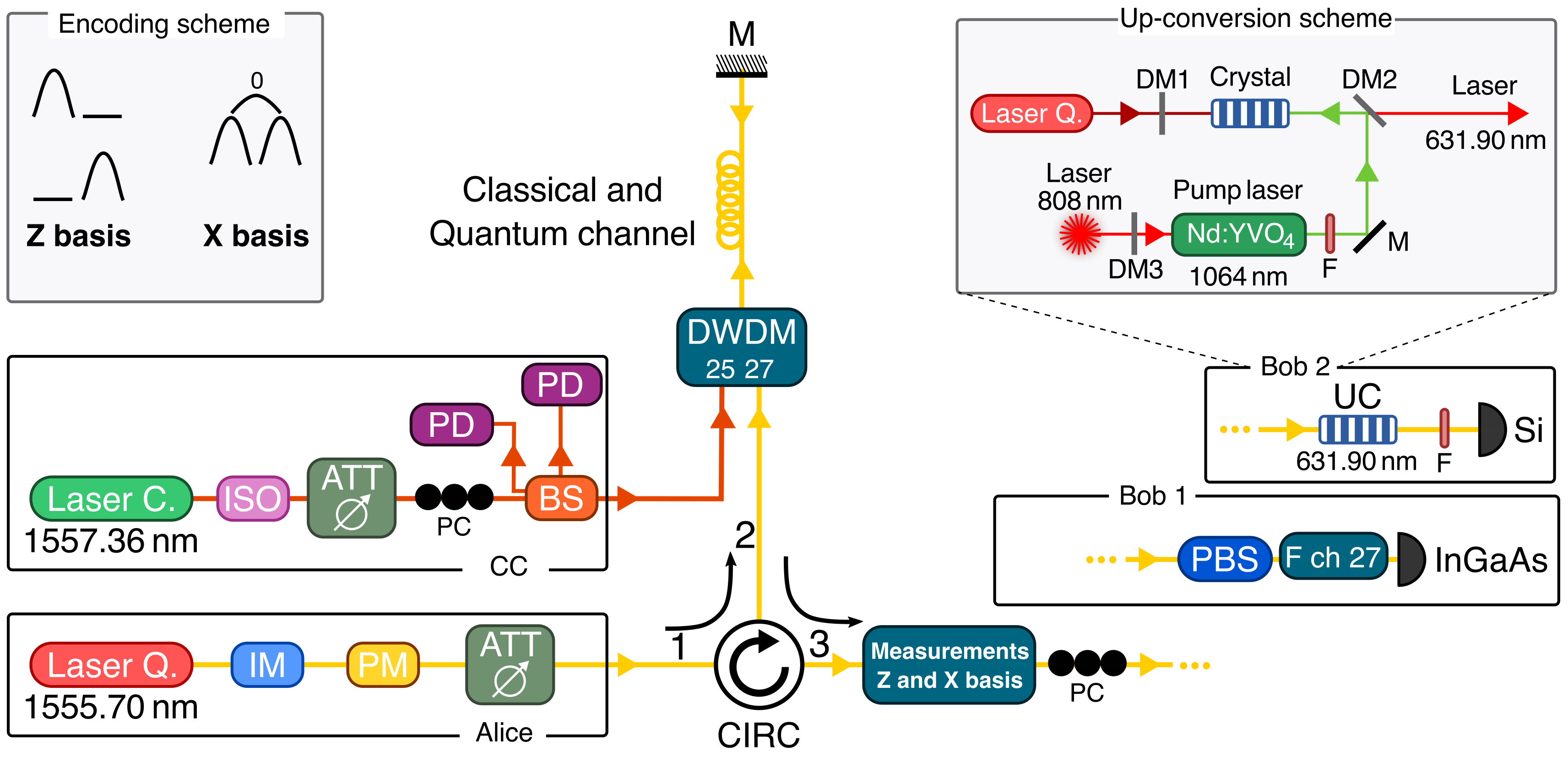}
\caption{\textbf{Setup of the experiment.} In the top left corner we report the encoding scheme adopted in the experiment. Legend: Laser C: classical laser, ISO: isolator, ATT: attenuator, PC: polarization controller, BS: beam splitter, PD: photodiode, DWDM: dense wavelength division multiplexing filter, M: fiber mirror, Laser Q: quantum laser, IM: intensity modulator, PM: phase-modulator, CIRC: circolator, PBS: polarization beam splitter, PC: polarization controller, F: filter, UC: up-conversion scheme (full setup in the Supplementary information), InGaAs: single-photon detector based on indium gallium arsenide, Si: silicon based single-photon detector.}
\label{fig:setup}
\end{figure*}

\subsection*{Protocol}
The QKD protocol used in this experiment is the three-state BB84 protocol with time-bin encoding~\cite{boaron2018secure,bacco2019fi}. The quantum states belonging to $\mathcal{Z}$ basis are used for the key generation process, and {\red} the $\mathcal{X}$ basis {\red{is}} used for the security check. More specifically, in the $\mathcal{Z}$ basis, one of the two time bins (early and late) is occupied by a weak coherent state, while the third state in the $\mathcal{X}$ basis is the equal superposition of the $\mathcal{Z}$ basis states with zero relative phase, as reported in the top-left {\red{corner}} of Figure \ref{fig:setup}. 
The security of {\red{the}} three-state protocol using finite-key analysis against general attacks has been demonstrated and combined with a very efficient one-decoy state scheme{\red{,}} in order to detect photon number splitting attacks \cite{Rusca2018_SecProofSimpleBB84,Rusca2018_FiniteKeyAnalysis}. The secret key rate (SKR) length ($ \ell $) per privacy amplification block is given by:
\begin{equation}
\begin{split}
    \ell \ \ \leq \ \  & D_0^{\mathcal{Z}} \ \ + \ \ D_1^{\mathcal{Z}} \Bigl[ 1-h\bigl( \phi_1^{\mathcal{Z}} \bigr) \Bigr] \ \ - \ \ \lambda_{EC} \ \\
    & \ \ - \ \ 6\log_2(19/\epsilon_{sec}) \ \ - \ \ \log_2(2/\epsilon_{corr})
\end{split}
\label{eq:rate}
\end{equation}
where $D_0^{\mathcal{Z}}$ and $D_1^{\mathcal{Z}}$ are the lower bounds of vacuum events and single-photon events in the $\mathcal{Z}$ basis, $h(\cdot)$ is the binary entropy function, $\phi_1^{\mathcal{Z}}$ is the upper bound on the phase error rate and $\lambda_{EC}$ is the number of bits that are publicly announced during error correction \cite{Rusca2018_SecProofSimpleBB84}. Finally, $\epsilon_{sec}$ and $\epsilon_{corr}$ are the secrecy and correctness parameters. In our computations we used a block size of $10^7$ bits and $\epsilon_{sec}=\epsilon_{corr}=10^{-9}$.

\section*{Experimental Setup}
In order to validate our hypothesis (i.e. up-conversion detection is more robust to spurious effects), we have tested the three-state time-bin protocol (as described above), exploiting weak coherent pulses and one-decoy method, combined with a classical laser co-propagating in the same fiber link.
As illustrated in Figure \ref{fig:setup}, the experimental setup consists of two optical transmitters (classical CC and quantum-Alice) and three receivers (one classical, and two quantum: Bob 1 and Bob 2) connected by a metropolitan dark-fiber link in a loop-back configuration. A fiber mirror is used to reflect the light back to the European Laboratory for Non-linear Spectroscopy (LENS), where the transmitters and receivers are located~\cite{bacco2019fi}. The installed fiber link is part of a fiber backbone provided by the Italian National Institute of Metrological Research (INRIM). 
In particular, we have used a QKD transmitter composed of a telecom laser at 1555.70 nm (channel 27 of the ITU-T 200 GHz grid) followed by two intensity modulators and a phase modulator~\cite{qti2020}. The two intensity modulators (IMs) are used for carving out the different time-bins and to implement the decoy state method. The phase modulator (PM) is used to set a random phase between different time-slots in order to assure the security against coherent attacks. The electrical outputs used to drive the IMs are provided by a field programmable gate array (FPGA), giving a state preparation rate of {\red} 595 MHz. Electrical pulse width is approximately $80$ ps, whereas the obtained optical pulse width is around $100$ ps. The PM is driven by a digital-to-analog converter which uses 8 bit to obtain $2^{8}-1$ different phase values. Furthermore, a pseudo random binary sequence of $2^{12}-1$ bits is used as a key generator, although a quantum random number generator should be adopted in a real implementation~\cite{Stevanov2000,Jennewein2000}. Subsequently to the IMs and PM, a variable optical attenuator (ATT) is employed to decrease the mean photon number per pulse to the quantum regime. 
The second transmitter (classical{\red{, CC in Figure \ref{fig:setup}}}) consists of a continuous wave (CW) laser at 1557.36 nm (channel 25 of the ITU-T 200 GHz grid) to emulate a classical communication link. The CC transmitter is then composed by an optical isolator, a variable optical attenuator and a beam-splitter (BS), which allows to monitor in real-time the optical power co-propagated in the fiber link. To be noted that although we did not encode any data {\red transmission} on the classical channel, a CW laser is perfectly able to emulate such a system for our proof-of-concept experiment.
The classical and  the quantum light are then combined by means of a DWDM filter, whose output is connected to the deployed fiber. After being reflected back by the fiber mirror, the light passes again through the same DWDM device, which separates the two different wavelengths (quantum and classical). 
The classical light is then measured at the output of the BS using a photodiode. {\red As illustrated in Figure \ref{fig:setup}, the experimental setup consists of two optical transmitters (classical CC and quantum-Alice) and three receivers (one classical, and two quantum: Bob 1 and Bob 2)}. On the contrary, the quantum light passes through the optical circulator (port 2) and is propagated through the exit (port 3). Here, we performed projective measurements in the $\mathcal{Z}$ and $\mathcal{X}$ bases. In particular, in the $\mathcal{Z}$ basis, the fiber is connected directly to one of the two detection schemes (Bob 1 or Bob 2), while and in the $\mathcal{X}$ basis the quantum states are sent to a free-space delay-line interferometer (with 4 dB loss) before reaching the detector, in order to monitor the relative phase within the two time-bins. The two single-photon detection schemes are, respectively, an up-conversion module~\cite{friis2019} (from 1555.70 nm to 631.90 nm), followed by a silicon-based photon counter from Micro Photon Devices~\cite{giudice2007}{\red{} and a free-running {\red fiber-based} InGaAs single-photon detector from ID Quantique (ID221). The detailed scheme of the up-conversion setup is reported in the Supplementary Material. Finally, a time-tagging unit, which is synchronized with the FPGA through an electrical clock signal, collects the measurement outputs from the detectors.\\ 
Although the total length of a round-trip in the fiber is about 40 km, with a overall loss of 21 dB, we have decided to use only a portion of this fiber as a quantum channel, in order to emulate different link configurations (i.e., 3, 5 and 8 dB of channel loss). In addition, the loss of the two detection systems, combined with the loss of the interferometer, has limited our ability to distribute the quantum signals over the entire link. It is important to notice that these restrictions are not limiting the overall idea, but the same principle can be tested in longer fiber links by accurately designing the QKD setup. 
Another important point to be mentioned is that although the actual quantum channel is shorter than the overall length of the fiber, the noise introduced by the the classical light is generated over the entire fiber {\red{link}}. In this condition, the noise level is overestimated with respect to the typical case of application.\\

\section*{{\red Up-conversion assisted detector}}
\begin{figure}[ht!]
\centering
\includegraphics[width=0.5\textwidth]{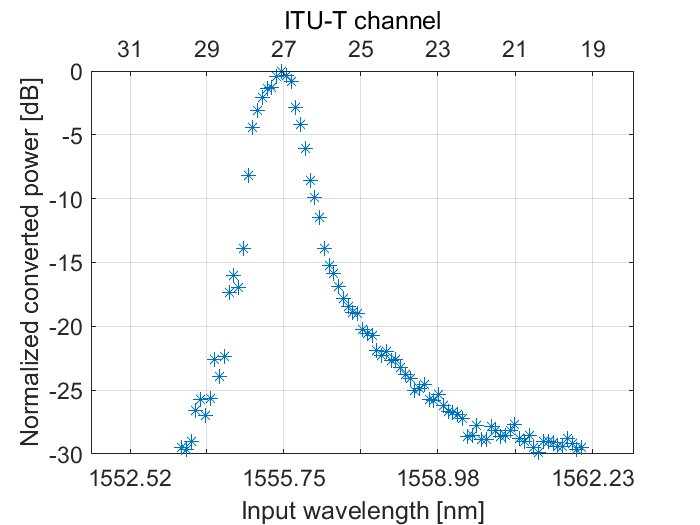}
\caption{{\red{\textbf{Phase matching profile.}}} {\red{Up-converted power (normalized) as a function of the input wavelength in the nonlinear crystal. The phase matching condition of the up-conversion process is optimized by 1555.70 nm of wavelength (channel 27 of the ITU-T grid), with a 3 dB bandwidth of 0.8 nm.
}}}
\label{fig:phase_matching2}
\end{figure}
{\red The up-conversion module is built with a high-finesse laser cavity (confined by mirrors DM1–DM3) in which a Nd:YVO4 crystal emitting at 1064 nm is pumped by an external laser diode at 808 nm \cite{friis2019upconversion}. Inside the cavity, a 40 mm nonlinear crystal is located in such a way that the intra-cavity field propagates in the direction of the poling. The quantum light at 1555.70 nm (corresponding to channel 27 of the ITU-T grid) is focused into the nonlinear crystal, where it is up-converted into 631.90 nm, which exits the cavity through DM2. More details on the up-conversion setup are reported in the Supplementary Material. In Figure \ref{fig:phase_matching2} we report the phase-matching profile of the up-conversion process, which acts as an intrinsic wavelength filter with a 3 dB bandwidth of 0.8 nm. In order to filter out the noisy nonlinear emission generated by the pump laser, four off-the-shelf optical filters (short-pass 650 nm, long-pass 600 nm, and two band-pass with 10 nm and 5 nm of bandwidth) are inserted before the free-space silicon-based single-photon counter (Micro Photon Devices, with a quantum efficiency of 40\% around 632 nm \cite{giudice2007high}). In our experiment, the overall efficiency of the up-conversion detector (including the conversion efficiency, filtering, coupling and silicon detector) is approximately 2\%, with an overall dark count rate of 11 kHz. With respect to our working point, the pump power at 1064 nm could be further increased to enhance the conversion efficiency (although, in this way, the dark count noise would be raised as well). On the contrary, the commercial InGaAs detector exhibits 20\% efficiency and 700 Hz of intrinsic dark count rate. However, even though the conversion process adversely affects the signal-to-noise ratio of the silicon detector, the up-conversion receiver still outperforms the InGaAs detector in terms of timing performances, thanks to the higher count rate (77 ns dead time) and ultra-low timing jitter (34 ps) of the Micro Photon Device module \cite{giudice2007high}. By contrast, the InGaAs detector requires a longer dead time (\SI{20}{\micro\second}) in order to avoid the high after-pulsing noise, which is further enhanced in the working condition of 20\% detection efficiency (condition which, on the other hand, is necessary to optimize the timing jitter to $\sim 200$ ps).
}

\section*{Noise evaluation and filtering}
In order to set the wavelength of the classical laser, we have decided to characterize the noise generated in our experimental setup, including the metropolitan fiber link and the DWDM device. Based on the assumption that the quantum laser was fixed at channel 27 due to the up-conversion module ({\red see Figure \ref{fig:phase_matching2}}), we have decided to evaluate the amount of spurious light scattered within the bandwidth of channel 27, as a function of the wavelength of the classical laser in input. The experimental setup for this characterization is reported in Figure \ref{fig:noise}a). By using the 200 GHz DWDM filter and a tunable laser source, we have tested one-by-one all the different ITU-T channels from 21 to 51 (at 0 dBm of input power) and we have measured the count rate at the output of channel 27 with an InGaAs single-photon detector. 
The normalized noise counts (after removing the averaged dark counts of the detector) are reported in Figure \ref{fig:noise}b). 
{\red It is clear that} channel 25 was found to introduce the highest noise counts in the quantum channel. The reason of this could be the specific configuration of our experimental setup, including device imperfections of the DWDM filter{\red{. A}}nyway, the noise counts were found to be independent from polarization. Based on this result, channel 25 was selected as the wavelength of the classical laser in our experiment, in order to test the QKD protocol under the worst condition {\red{of noise}}. \\ 
Furthermore, since the up-conversion unit is both polarization dependent {\red (due to the nonlinear process)} and wavelength dependent {\red (see Figure \ref{fig:phase_matching2})}, we have decided to include analogous advantages also in the InGaAs receiver, as depicted in Bob 1 setup{\red{, by adding off-the-shelf devices in front of the InGaAs single-photon detector}}. Specifically, we {\red employed a polarizing beam splitter (PBS) and} a 100 GHz band-pass filter of channel 27{\red, exhibiting 0.64 nm of 3 dB bandwidth and 80 dB of extinction ratio between channels 25 and 27}. In addition, {\red we decided to test both receivers in the condition of orthogonal polarization directions of quantum signals and residual classical noise incoming at the QKD detector. To do so,} we have included a polarization controller {\red (PC)} in the CC transmitter, and we have tuned it in order to minimize the amount of noise counts into both detection systems. In Bob 1 setup, {\red used to filter out the noise from classical light before the InGaAs detector{\red, with an overall insertion loss of 6 dB}. In Bob 2 setup, polarization {\red and wavelength} filtering {\red are} provided inherently by the up-conversion process. {\red Another PC} is put in front of both receivers, in order to align the polarization of quantum light with the filtered direction.\\

\begin{figure}[ht!]
\centering
\includegraphics[width=0.5\textwidth]{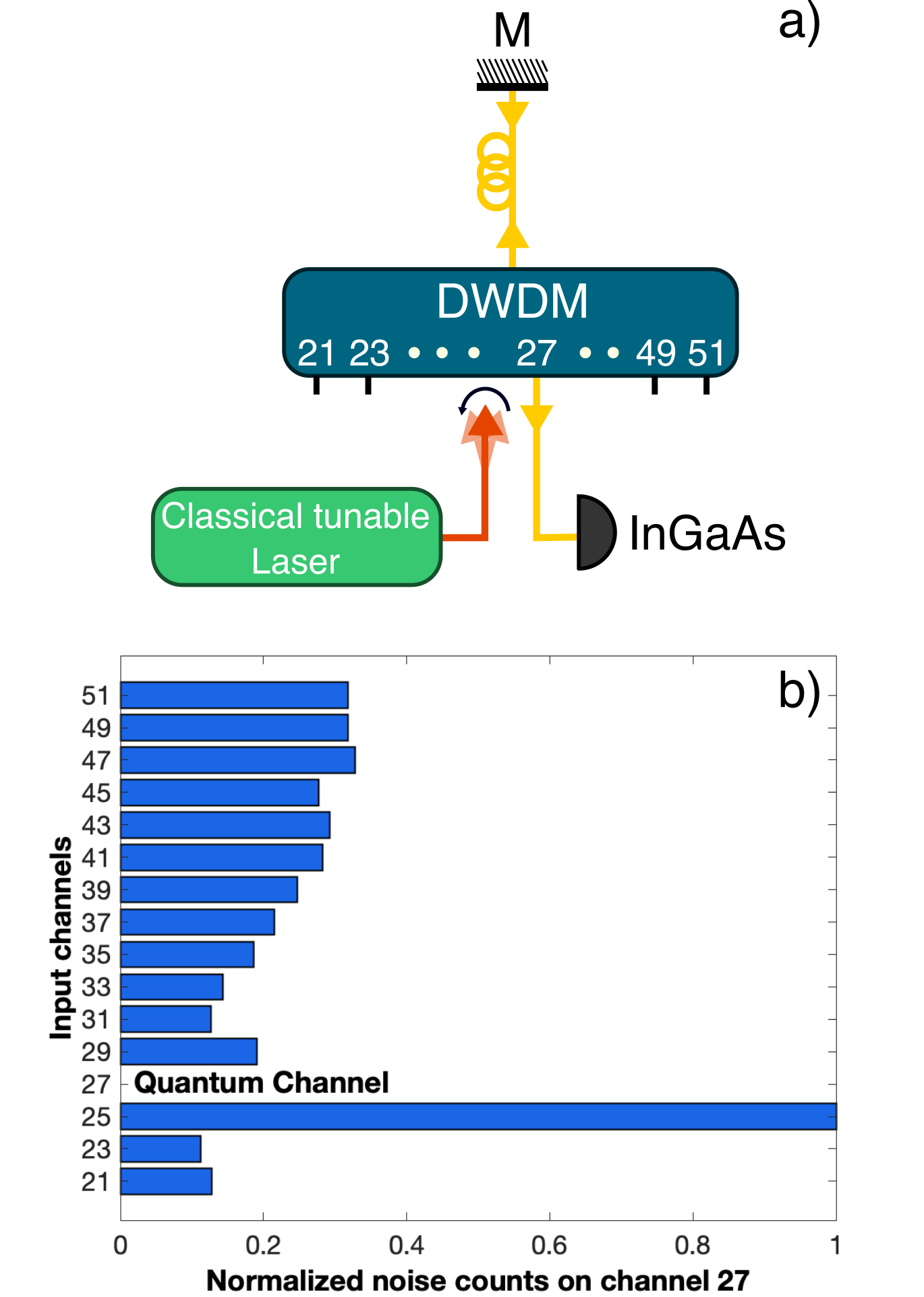}
\caption{\textbf{Noise characterization of the experimental setup.} Noise counts within the bandwidth of channel 27 are reported as a function of the different wavelength of a tunable laser entering the DWDM device.}
\label{fig:noise}
\end{figure}

To be noted that the polarization drift in a deployed fiber is {\red slow with respect to the typical QKD acquisition time}, as demonstrated by the long-term acquisitions reported in previous works \cite{bacco2019fi,wengerowsky2019entanglement}.\\
Finally, temporal filtering of the time-bin windows is used to post-select the acquired clicks from both detectors.

\section*{Results}
In order to test the QKD protocol, the experimental parameters (such as the mean photon number of signal and decoy states, and their probability of preparation) were mathematically optimized for the two types of receiver, in order to maximize the secret key rate achievable at the different channel lengths. The detailed values that we set are reported in the Supplementary Material. 
We experimentally measured the quantum bit error rate (QBER) exhibited by the two receivers in the two mutually unbiased bases, for different classical power levels at the DWDM input, ranging from -20 dBm to -8 dBm. The results are reported in Figure \ref{fig:comparison}a), \ref{fig:comparison}b), \ref{fig:comparison}d), and \ref{fig:comparison}e). \\
After the collection of the data, by using Eq.~\eqref{eq:rate} we estimated the secret key rate achievable with the two receivers, that is reported in Figure \ref{fig:comparison}c) and \ref{fig:comparison}f). 
The black line represents the numerical simulation of the QKD performance in the back-to-back configuration, i.e., without using the metropolitan fiber as transmission channel and without co-propagating the classical laser. The experimental data acquired in this configuration are given by the filled {\red circles} in Figure \ref{fig:comparison}.

\begin{figure*}[ht!]
\centering
\includegraphics[width=0.99\textwidth]{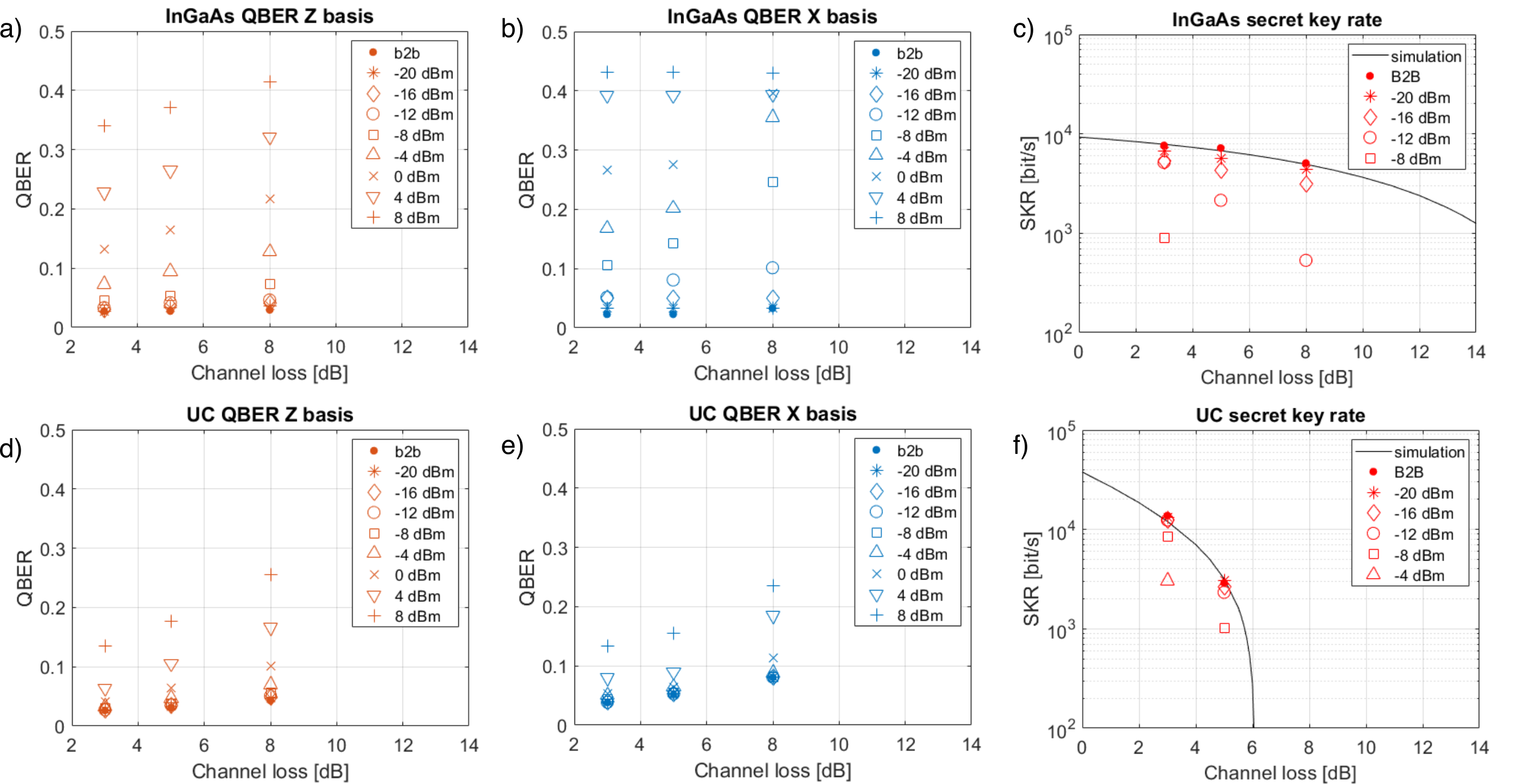}
\caption{\textbf{Experimental results.} Each point represents the averaged quantum bit error rate and secret key rate, as a function of channel loss, experimentally acquired in the back-to-back configuration (B2B) and for the different power levels of the classical laser entering in the DWDM device, ranging from -20 dBm to -8 dBm. 
\textbf{a), b), e), f)}: quantum bit error rate in the $\mathcal{Z}$ and  $\mathcal{X}$ basis, measured with the InGaAs detector and the up-conversion unit, respectively. \textbf{c), d)}: secret key rate achievable by the two different QKD receivers. 
}
\label{fig:comparison}
\end{figure*}

\begin{figure}[ht!]
\centering
\includegraphics[width=0.45\textwidth]{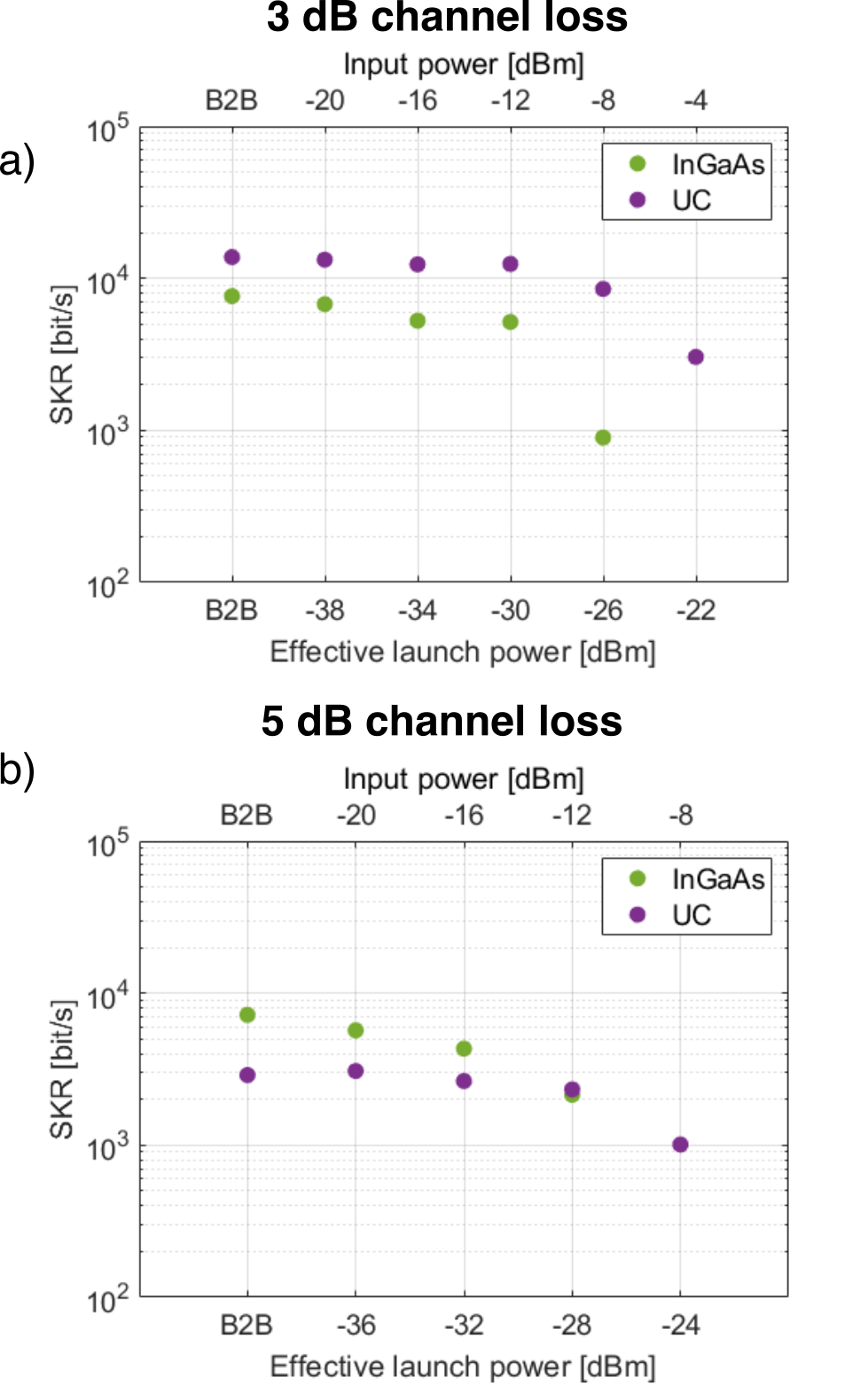}
\caption{\textbf{Secret key rate as a function of the classical launch power}. Here are reported the experimental data from Figure \ref{fig:comparison}c) and \ref{fig:comparison}f), as a function of the classical launch power that is actually injected into the quantum channel, a) at 3 dB channel loss and b) at 5 dB loss.
}
\label{fig:pote_rate}
\end{figure}

\section*{Discussion}
In this proof-of-concept experiment, we tested the ability of the up-conversion-assisted QKD to tolerate more noise in the quantum channel, as compared to a standard detector. 
The final figure of merit is represented by the secret key rate, which is a fundamental parameter in the telecommunication system (the higher the key generation rate, the faster is the key refresh rate for data encryption). 
More specifically, we reported in Figure \ref{fig:comparison}c) and \ref{fig:comparison}d) the amount of secret key (bit/s) as a function of the channel losses for different classical power levels. It is quite clear that, although the up-conversion scheme is inherently affected by high dark count rate (about 11 kHz) and low overall detection efficiency (around 2\%), which also limit the transmission distance of QKD, the amount of tolerable classical power is 4 dB larger for this receiver, both at 3 and 5 dB channel loss. As an example, by considering 5 dB channel loss, Bob 1 can tolerate up to -12 dBm of input power, as shown in Figure \ref{fig:pote_rate}b). Since the 5-dB quantum channel is only a portion of the overall 21-dB attenuation in the loop-back fiber link, the actual launch power of classical light in the quantum channel is -28 dBm, corresponding to {\red{-}}12 dBm of power at the DWDM input. On the other hand, at the same channel loss of 5 dB, the up-conversion assisted Bob 2 can tolerate up to -8 dBm of input power (corresponding to -24 dBm of launch power in the quantum channel). An overall gain of 4 dB in telecommunication systems is a big step since it makes possible at least to double the amount of data transmitted into the single mode fiber, in order to enhance the signal-to-noise ratio, thus decreasing the amount of errors. The same behaviour is reported in the case of 3 dB quantum channel, as reported in Figure \ref{fig:pote_rate}a). This figure gives a clear indication that the up-conversion receiver for QKD can tolerate more optical power in the quantum channel, at least for short-link configurations.\\
The advantage of the up-conversion unit, as compared to the standard QKD receiver, is the intrinsic filter in polarization and wavelength provided by the non-linear crystal, that is combined also with the silicon-based single-photon detector, exhibiting far better timing performances than the InGaAs detector. In particular, the ultra-low timing jitter (35 ps) provided by the silicon detector, allows for a more efficient post-selection filter of the time-bin windows, thus reducing the impact of dark and noise counts. 
In addition, the shorter dead time of silicon detector (77 ns) enables a higher click rate than the InGaAs detector, which is typically limited by saturation {\red effect}. {\red} 
Conversely, the up-conversion detector exhibits very high intrinsic noise, due to the multiple nonlinear effects and scattering generated by the intense laser pump at 1064 nm. {\red {\red With this setup}, our up-conversion receiver} can work only for short-distance QKD, below 6 dB channel loss, as shown in Figure \ref{fig:comparison}f). 
Nonetheless, state-of-the art systems for up-conversion detectors of single-photon signals in the C band, have demonstrated a very high overall efficiency above 30\%, with a dark count rate as low as 100 Hz, thus enabling quantum communication up to 45 dB of channel loss \cite{shentu2013ultralow,ma2018upconversion,yao2020optimizing}. 
Furthermore, it is important to notice that although we have used a bulky {\red and custom home-made} system {\red(which requires the pre-alignment of the free-space silicon detector)}, the nonlinear crystal could be integrated in photonic platforms for a more efficient and stable solution \cite{adcock2020review}. 

Regarding the InGaAs detector, we would like to stress once again that we have used a commercial device, equipped with an off-the-shelf band-pass filter {\red of 100 GHz bandwidth,} combined with a polarization beam splitter for polarization filtering. Ad-hoc components could be used and designed for improving the signal-to-noise ratio of the InGaAs detector, but our idea was to compare the up-conversion unit, which intrinsically offers a filter action, with a standard commercial quantum receiver. 

Finally, another important point to be considered in the key generation process, is the amount of time required for the key establishment. 
In particular, the acquisition time necessary to collect a block size of raw key bits depends on the click rate of the detector, thus the acquisition is expected to be faster with the up-conversion receiver. In our experiment, to collect a block size of $10^7$ bits at 5 dB channel loss, the acquisition takes about 3 minutes with the up-conversion unit, and 10 minutes with the standard receiver. 

In conclusion, we have demonstrated that by exploiting an up-conversion unit, in a quantum key distribution scheme, is possible to tolerate higher classical power in the optical channel compared to a standard InGaAs detector{\red{ equipped with off-the-shelf filtering devices}}. This proof-of-concept experiment represents a pivotal step towards the full integration of quantum and classical light within the same infrastructure. Furthermore, we believe it can pave the way to wider quantum applications in the deployed infrastructure.

\section*{Acknowledgements}
\vspace{-0.25cm}
The authors would like to thank D. Rusca and B. Da Lio for the fruitful discussion.

\section*{Funding}
\vspace{-0.25cm}
This work is supported by the Center of Excellence SPOC - Silicon Photonics for Optical Communications (ref DNRF123), by the EraNET Cofund Initiatives QuantERA within the European Union’s Horizon 2020 research and innovation program grant agreement No.731473 (project SQUARE) and by the NATO Science for Peace and Security program under Grant No. G5485.

\subsection*{Competing financial interests}
The authors declare that there are no competing interests.

\subsection*{Data availability}
The data that support the findings of this study are available on request from the corresponding author. The data are not publicly available due to privacy or ethical restrictions.

\cleardoublepage

\begin{thebibliography}{10}
\providecommand{\url}[1]{#1}
\csname url@samestyle\endcsname
\providecommand{\newblock}{\relax}
\providecommand{\bibinfo}[2]{#2}
\providecommand{\BIBentrySTDinterwordspacing}{\spaceskip=0pt\relax}
\providecommand{\BIBentryALTinterwordstretchfactor}{4}
\providecommand{\BIBentryALTinterwordspacing}{\spaceskip=\fontdimen2\font plus
\BIBentryALTinterwordstretchfactor\fontdimen3\font minus
  \fontdimen4\font\relax}
\providecommand{\BIBforeignlanguage}[2]{{%
\expandafter\ifx\csname l@#1\endcsname\relax
\typeout{** WARNING: IEEEtran.bst: No hyphenation pattern has been}%
\typeout{** loaded for the language `#1'. Using the pattern for}%
\typeout{** the default language instead.}%
\else
\language=\csname l@#1\endcsname
\fi
#2}}
\providecommand{\BIBdecl}{\relax}
\BIBdecl

\bibitem{pirandola2020advances}
S.~Pirandola, U.~L. Andersen, L.~Banchi, M.~Berta, D.~Bunandar, R.~Colbeck,
  D.~Englund, T.~Gehring, C.~Lupo, C.~Ottaviani \emph{et~al.}, ``Advances in
  quantum cryptography,'' \emph{Advances in Optics and Photonics}, vol.~12,
  no.~4, pp. 1012--1236, 2020.

\bibitem{xu2020secure}
F.~Xu, X.~Ma, Q.~Zhang, H.-K. Lo, and J.-W. Pan, ``Secure quantum key
  distribution with realistic devices,'' \emph{Reviews of Modern Physics},
  vol.~92, no.~2, p. 025002, 2020.

\bibitem{shimizu2014}
K.~Shimizu, T.~Honjo, M.~Fujiwara, T.~Ito, K.~Tamaki, S.~Miki, T.~Yamashita,
  H.~Terai, Z.~Wang, and M.~Sasaki, ``Performance of long-distance quantum key
  distribution over 90-km optical links installed in a field environment of
  {T}okyo metropolitan area,'' \emph{Journal of Lightwave Technology}, vol.~32,
  no.~1, pp. 141--151, 2014.

\bibitem{tang2016}
Y.-L. Tang, H.-L. Yin, Q.~Zhao, H.~Liu, X.-X. Sun, M.-Q. Huang, W.-J. Zhang,
  S.-J. Chen, L.~Zhang, L.-X. You \emph{et~al.},
  ``Measurement-device-independent quantum key distribution over untrustful
  metropolitan network,'' \emph{Physical Review X}, vol.~6, no.~1, p. 011024,
  2016.

\bibitem{collins2016}
R.~J. Collins, R.~Amiri, M.~Fujiwara, T.~Honjo, K.~Shimizu, K.~Tamaki,
  M.~Takeoka, E.~Andersson, G.~S. Buller, and M.~Sasaki, ``Experimental
  transmission of quantum digital signatures over 90 km of installed optical
  fiber using a differential phase shift quantum key distribution system,''
  \emph{Optics letters}, vol.~41, no.~21, pp. 4883--4886, 2016.

\bibitem{zhang2017}
Y.~Zhang, Z.~Li, Z.~Chen, C.~Weedbrook, Y.~Zhao, X.~Wang, Y.~Huang, C.~Xu,
  X.~Zhang, Z.~Wang \emph{et~al.}, ``Continuous-variable {Q}{K}{D} over 50 km
  commercial fiber,'' \emph{Quantum Science and Technology}, vol.~4, no.~3, p.
  035006, 2019.

\bibitem{bacco2019boosting}
D.~Bacco, B.~Da~Lio, D.~Cozzolino, F.~Da~Ros, X.~Guo, Y.~Ding, Y.~Sasaki,
  K.~Aikawa, S.~Miki, H.~Terai \emph{et~al.}, ``Boosting the secret key rate in
  a shared quantum and classical fibre communication system,''
  \emph{Communications Physics}, vol.~2, no.~1, pp. 1--8, 2019.

\bibitem{avesani2020}
M.~Avesani, L.~Calderaro, G.~Foletto, C.~Agnesi, F.~Picciariello,
  F.~Santagiustina, A.~Scriminich, A.~Stanco, F.~Vedovato, M.~Zahidy
  \emph{et~al.}, ``Resource-effective quantum key distribution: a field-trial
  in {P}adua city center,'' \emph{arXiv preprint:2012.08457}, 2020.

\bibitem{bacco2019fi}
D.~Bacco, I.~Vagniluca, B.~Da~Lio, N.~Biagi, A.~Della~Frera, D.~Calonico,
  C.~Toninelli, F.~S. Cataliotti, M.~Bellini, L.~K. Oxenl{\o}we \emph{et~al.},
  ``Field trial of a three-state quantum key distribution scheme in the
  {F}lorence metropolitan area,'' \emph{EPJ Quantum Technology}, vol.~6, no.~1,
  p.~5, 2019.

\bibitem{cozzolino2019high}
D.~Cozzolino, B.~Da~Lio, D.~Bacco, and L.~K. Oxenl{\o}we, ``High-dimensional
  quantum communication: Benefits, progress, and future challenges,''
  \emph{Advanced Quantum Technologies}, vol.~2, no.~12, p. 1900038, 2019.

\bibitem{mower2013high}
J.~Mower, Z.~Zhang, P.~Desjardins, C.~Lee, J.~H. Shapiro, and D.~Englund,
  ``High-dimensional quantum key distribution using dispersive optics,''
  \emph{Physical Review A}, vol.~87, no.~6, p. 062322, 2013.

\bibitem{bunandar2015practical}
D.~Bunandar, Z.~Zhang, J.~H. Shapiro, and D.~R. Englund, ``Practical
  high-dimensional quantum key distribution with decoy states,'' \emph{Physical
  Review A}, vol.~91, no.~2, p. 022336, 2015.

\bibitem{Steinlechner2017}
F.~Steinlechner, S.~Ecker, M.~Fink, B.~Liu, J.~Bavaresco, M.~Huber, T.~Scheidl,
  and R.~Ursin, ``Distribution of high-dimensional entanglement via an
  intra-city free-space link,'' \emph{Nature Communication}, vol.~8, no. 15971,
  2017.

\bibitem{Martin2017}
A.~Martin, T.~Guerreiro, A.~Tiranov, S.~Designolle, F.~Fr\"owis, N.~Brunner,
  M.~Huber, and N.~Gisin, ``Quantifying photonic high-dimensional
  entanglement,'' \emph{Physical Review Letters}, vol. 118, p. 110501, 2017.

\bibitem{Wang2018}
J.~Wang, S.~Paesani, Y.~Ding, R.~Santagati, P.~Skrzypczyk, A.~Salavrakos,
  J.~Tura, R.~Augusiak, L.~Man{\v c}inska, D.~Bacco, D.~Bonneau, J.~W.
  Silverstone, Q.~Gong, A.~Ac{\'\i}n, K.~Rottwitt, L.~K. Oxenl{\o}we, J.~L.
  O{\textquoteright}Brien, A.~Laing, and M.~G. Thompson, ``Multidimensional
  quantum entanglement with large-scale integrated optics,'' \emph{Science},
  vol. 360, no. 6386, pp. 285--291, 2018.

\bibitem{Ding2017}
Y.~Ding, D.~Bacco, K.~Dalgaard, X.~Cai, X.~Zhou, K.~Rottwitt, and L.~K.
  Oxenl{\o}we, ``High-dimensional quantum key distribution based on multicore
  fiber using silicon photonic integrated circuits,'' \emph{npj Quantum
  Information}, vol.~3, no.~1, p.~25, 2017.

\bibitem{islam2017provably}
N.~T. Islam, C.~C.~W. Lim, C.~Cahall, J.~Kim, and D.~J. Gauthier, ``Provably
  secure and high-rate quantum key distribution with time-bin qudits,''
  \emph{Science advances}, vol.~3, no.~11, p. e1701491, 2017.

\bibitem{vagniluca2020}
I.~Vagniluca, B.~Da~Lio, D.~Rusca, D.~Cozzolino, Y.~Ding, H.~Zbinden,
  A.~Zavatta, L.~K. Oxenl\o{}we, and D.~Bacco, ``Efficient time-bin encoding
  for practical high-dimensional quantum key distribution,'' \emph{Physical
  Review Applied}, vol.~14, p. 014051, 2020.

\bibitem{tanaka2008}
A.~Tanaka, M.~Fujiwara, S.~W. Nam, Y.~Nambu, S.~Takahashi, W.~Maeda, K.-i.
  Yoshino, S.~Miki, B.~Baek, Z.~Wang \emph{et~al.}, ``Ultra fast quantum key
  distribution over a 97 km installed telecom fiber with wavelength division
  multiplexing clock synchronization,'' \emph{Optics express}, vol.~16, no.~15,
  pp. 11\,354--11\,360, 2008.

\bibitem{choi2014}
I.~Choi, Y.~R. Zhou, J.~F. Dynes, Z.~Yuan, A.~Klar, A.~Sharpe, A.~Plews,
  M.~Lucamarini, C.~Radig, J.~Neubert \emph{et~al.}, ``Field trial of a quantum
  secured 10 {G}b/s {DWDM} transmission system over a single installed fiber,''
  \emph{Optics express}, vol.~22, no.~19, pp. 23\,121--23\,128, 2014.

\bibitem{mao2018}
Y.~Mao, B.-X. Wang, C.~Zhao, G.~Wang, R.~Wang, H.~Wang, F.~Zhou, J.~Nie,
  Q.~Chen, Y.~Zhao \emph{et~al.}, ``Integrating quantum key distribution with
  classical communications in backbone fiber network,'' \emph{Optics express},
  vol.~26, no.~5, pp. 6010--6020, 2018.

\bibitem{gustavo2020}
G.~B. Xavier and G.~Lima, ``Quantum information processing with space-division
  multiplexing optical fibres,'' \emph{Communications Physics}, vol.~3, no.~1,
  pp. 1--11, 2020.

\bibitem{peters2009dense}
N.~Peters, P.~Toliver, T.~Chapuran, R.~Runser, S.~McNown, C.~Peterson,
  D.~Rosenberg, N.~Dallmann, R.~Hughes, K.~McCabe \emph{et~al.}, ``Dense
  wavelength multiplexing of 1550 nm qkd with strong classical channels in
  reconfigurable networking environments,'' \emph{New Journal of physics},
  vol.~11, no.~4, p. 045012, 2009.

\bibitem{eraerds2010}
P.~Eraerds, N.~Walenta, M.~Legr{\'e}, N.~Gisin, and H.~Zbinden, ``Quantum key
  distribution and 1 gbps data encryption over a single fibre,'' \emph{New
  Journal of Physics}, vol.~12, no.~6, p. 063027, 2010.

\bibitem{qi2010feasibility}
B.~Qi, W.~Zhu, L.~Qian, and H.-K. Lo, ``Feasibility of quantum key distribution
  through a dense wavelength division multiplexing network,'' \emph{New Journal
  of Physics}, vol.~12, no.~10, p. 103042, 2010.

\bibitem{kumar2015coexistence}
R.~Kumar, H.~Qin, and R.~All{\'e}aume, ``Coexistence of continuous variable qkd
  with intense dwdm classical channels,'' \emph{New Journal of Physics},
  vol.~17, no.~4, p. 043027, 2015.

\bibitem{eriksson2019wavelength}
T.~A. Eriksson, T.~Hirano, B.~J. Puttnam, G.~Rademacher, R.~S. Lu{\'\i}s,
  M.~Fujiwara, R.~Namiki, Y.~Awaji, M.~Takeoka, N.~Wada \emph{et~al.},
  ``Wavelength division multiplexing of continuous variable quantum key
  distribution and 18.3 tbit/s data channels,'' \emph{Communications Physics},
  vol.~2, no.~1, pp. 1--8, 2019.

\bibitem{boaron2018secure}
A.~Boaron, G.~Boso, D.~Rusca, C.~Vulliez, C.~Autebert, M.~Caloz, M.~Perrenoud,
  G.~Gras, F.~Bussi{\`e}res, M.-J. Li \emph{et~al.}, ``Secure quantum key
  distribution over 421 km of optical fiber,'' \emph{Physical Review Letters},
  vol. 121, no.~19, p. 190502, 2018.

\bibitem{Rusca2018_SecProofSimpleBB84}
D.~Rusca, A.~Boaron, M.~Curty, A.~Martin, and H.~Zbinden, ``Security proof for
  a simplified {B}ennett-{B}rassard 1984 quantum-key-distribution protocol,''
  \emph{Physical Review A}, vol.~98, no.~5, p. 052336, 2018.

\bibitem{Rusca2018_FiniteKeyAnalysis}
D.~Rusca, A.~Boaron, F.~Gr{\"u}nenfelder, A.~Martin, and H.~Zbinden,
  ``Finite-key analysis for the 1-decoy state {QKD} protocol,'' \emph{Applied
  Physics Letters}, vol. 112, no.~17, p. 171104, 2018.

\bibitem{qti2020}
``Qti s.r.l.'' \emph{qticompany.com}.

\bibitem{Stevanov2000}
A.~Stefanov, N.~Gisin, O.~Guinnard, L.~Guinnard, and H.~Zbinden, ``Optical
  quantum random number generator,'' \emph{Journal of Modern Optics}, vol.~47,
  no.~4, pp. 595--598, 2000.

\bibitem{Jennewein2000}
T.~Jennewein, U.~Achleitner, G.~Weihs, H.~Weinfurter, and A.~Zeilinger, ``A
  fast and compact quantum random number generator,'' \emph{Review of
  Scientific Instruments}, vol.~71, no.~4, pp. 1675--1680, 2000.

\bibitem{friis2019}
S.~M. Friis and L.~H{\o}gstedt, ``Upconversion-based mid-infrared spectrometer
  using intra-cavity linbo 3 crystals with chirped poling structure,''
  \emph{Optics letters}, vol.~44, no.~17, pp. 4231--4234, 2019.

\bibitem{giudice2007}
A.~Giudice, M.~Ghioni, R.~Biasi, F.~Zappa, S.~Cova, P.~Maccagnani, and
  A.~Gulinatti, ``High-rate photon counting and picosecond timing with
  silicon-spad based compact detector modules,'' \emph{Journal of Modern
  Optics}, vol.~54, no. 2-3, pp. 225--237, 2007.

\bibitem{friis2019upconversion}
S.~M. Friis and L.~H{\o}gstedt, ``Upconversion-based mid-infrared spectrometer
  using intra-cavity linbo 3 crystals with chirped poling structure,''
  \emph{Optics letters}, vol.~44, no.~17, pp. 4231--4234, 2019.

\bibitem{giudice2007high}
A.~Giudice, M.~Ghioni, R.~Biasi, F.~Zappa, S.~Cova, P.~Maccagnani, and
  A.~Gulinatti, ``High-rate photon counting and picosecond timing with
  silicon-spad based compact detector modules,'' \emph{Journal of Modern
  Optics}, vol.~54, no. 2-3, pp. 225--237, 2007.

\bibitem{wengerowsky2019entanglement}
S.~Wengerowsky, S.~K. Joshi, F.~Steinlechner, J.~R. Zichi, S.~M. Dobrovolskiy,
  R.~van~der Molen, J.~W. Los, V.~Zwiller, M.~A. Versteegh, A.~Mura
  \emph{et~al.}, ``Entanglement distribution over a 96-km-long submarine
  optical fiber,'' \emph{Proceedings of the National Academy of Sciences}, vol.
  116, no.~14, pp. 6684--6688, 2019.

\bibitem{shentu2013ultralow}
G.-L. Shentu, J.~S. Pelc, X.-D. Wang, Q.-C. Sun, M.-Y. Zheng, M.~Fejer,
  Q.~Zhang, and J.-W. Pan, ``Ultralow noise up-conversion detector and
  spectrometer for the telecom band,'' \emph{Optics express}, vol.~21, no.~12,
  pp. 13\,986--13\,991, 2013.

\bibitem{ma2018upconversion}
F.~Ma, L.-Y. Liang, J.-P. Chen, Y.~Gao, M.-Y. Zheng, X.-P. Xie, H.~Liu,
  Q.~Zhang, and J.-W. Pan, ``Upconversion single-photon detectors based on
  integrated periodically poled lithium niobate waveguides,'' \emph{JOSA B},
  vol.~35, no.~9, pp. 2096--2101, 2018.

\bibitem{yao2020optimizing}
N.~Yao, Q.~Yao, X.-P. Xie, Y.~Liu, P.~Xu, W.~Fang, M.-Y. Zheng, J.~Fan,
  Q.~Zhang, L.~Tong \emph{et~al.}, ``Optimizing up-conversion single-photon
  detectors for quantum key distribution,'' \emph{Optics Express}, vol.~28,
  no.~17, pp. 25\,123--25\,133, 2020.

\bibitem{adcock2020review}
J.~C. Adcock, J.~Bao, Y.~Chi, X.~Chen, D.~Bacco, Q.~Gong, L.~K. Oxenl{\o}we,
  J.~Wang, and Y.~Ding, ``Advances in silicon quantum photonics,'' \emph{IEEE
  Journal of Selected Topics in Quantum Electronics}, vol.~27, no.~2, pp.
  1--24, 2020.

\end{thebibliography}

\begin{thebibliography}{}

\bibitem{friis2019} Søren MM Friis, Lasse Høgstedt, "Upconversion-based mid-infrared spectrometer using intra-cavity LiNbO 3 crystals with chirped poling structure" \textit{Optics letters}, vol. 44.17, p. 4231-4234, 2019.

\bibitem{giudice2007} A. Giudice, M. Ghioni, R. Biasi, F. Zappa, S. D. Cova, P. Maccagnani, and A. Gulinatti, "High-rate photon counting and picosecond timing with silicon-SPAD based compact detector modules" \textit{Journal of Modern Optics}, vol. 54, no. 2, pp. 225–237, 2007
\end{thebibliography}


\newpage \onecolumn

\begin{center}{\title{\Large{\textbf{Supplementary Information}}\\ \vspace{6pt} \large{\textbf{Towards fully-fledged quantum and classical communication over deployed fiber with up-conversion module}}}} 

\vspace{4pt}\author{Davide Bacco$^{1* \dagger}$, Ilaria Vagniluca $^{2,3 \dagger}$, Daniele Cozzolino$^{1}$, Søren M. M. Friis$^{2}$,Lasse Høgstedt$^{2}$,Andrea Giudice$^{3}$,Davide Calonico$^{6}$, Francesco Saverio Cataliotti$^{3,7,8}$,\\ Karsten Rottwitt$^{1}$, Alessandro Zavatta$^{3,7,8}$}

\vspace{12pt}\textit{\small{
$^\textrm{\textit{1}}$Center for Silicon Photonics for Optical Communication (SPOC), Department of Photonics Engineering, Technical University of Denmark, 2800 Kgs. Lyngby, Denmark.\\
$^\textrm{\textit{2}}$ Department of Physics “Ettore Pancini", University of Naples “Federico II", 80126 Naples, IT\\
$^\textrm{\textit{3}}$CNR - Istituto Nazionale di Ottica (CNR-INO), Largo E. Fermi, 6 - 50125 Firenze, Italy.\\
$^\textrm{\textit{4}}$NLIR ApS, Hirsemarken 1 1st floor, 3520 Farum, Denmark.\\
$^\textrm{\textit{5}}$Micro Photon Devices S.r.l., via Antonio Stradivari 4, 39100 Bolzano, Italy\\
$^\textrm{\textit{6}}$ I.N.Ri.M. Istituto Nazionale di Ricerca Metrologica, Torino,Italy\\
$^\textrm{\textit{7}}$LENS and Dipartimento di Fisica e Astronomia, Università di Firenze, Via G. Sansone, 1 - 50019 Sesto Fiorentino, Italy.\\
$^\textrm{\textit{8}}$ QTI SRL, Largo Enrico Fermi, 6 - 50125 Firenze, Italy.\\
$\dagger$ These authors contributed equally to this work\\
* dabac@fotonik.dtu.dk
}}

\end{center}

\setcounter{section}{0}
\renewcommand{\thesection}{Supplementary Note~\arabic{section}:}
\setcounter{figure}{0}
\renewcommand{\figurename}{Supplementary Figure}
\setcounter{table}{0}
\renewcommand{\tablename}{Supplementary Table}
\setcounter{equation}{0}
\renewcommand\theequation{S\arabic{equation}}
\renewcommand{\refname}{Supplementary References}

\section*{Up-conversion scheme}
One of the two receivers presented in the main text, is realized with a frequency up-conversion unit which is depicted in Figure \ref{fig:up_conversion_scheme}. The up-conversion module is built with a high-finesse laser cavity confined by mirrors DM1–DM3 in which the gain medium is a Nd:YVO4 crystal, emitting at 1064 nm and pumped by an external laser diode at 808 nm \cite{friis2019}. The filter (F) inside the cavity consists of four band-stop mirrors centered at 1064 nm that remove at eight orders of magnitude of residual laser diode light and fluorescence generated from the crystal. More specifically, one mirror is concave with a focal length of 75 mm in order to make the entire cavity stable and to reduce the spot size of the incoming beam size of 180 $\mu$m inside the nonlinear crystal. The three mirrors DM1–DM3 are all high-reflection (HR) coated for 1064 nm and additionally the first mirror (DM1) is anti-reflection (AR) coated for mid-infrared wavelengths (MIR). In addition, the DM2 is AR coated for near-visible light up to 1000 nm, and DM3 is AR coated for 808 nm. 
Inside the cavity there is a 40 mm long nonlinear crystal (PPLN) which is located such that the intra-cavity field propagates in the direction of the poling. The nonlinear crystal is AR coated for 1064 nm on both end facets and AR coated for MIR wavelengths (left side on Figure\ref{fig:up_conversion_scheme}) and AR coated for the up-converted wavelengths (the right side Figure\ref{fig:up_conversion_scheme}). The quantum light at 1555.70 nm (channel 27 of the ITU-T grid) is focused into the non-linear crystal where it is up-converted to the visible regime, and it exits the cavity through DM2 at 631.90 nm. 
Furthermore, in order to increase the signal-to-noise ratio and to limit the different noise generated in the up-conversion unit, we have used four off-the-shelf optical filters (low-pass, high-pass filter, band-pass 5 nm and band-pass 10 nm). After the filters the light is collected by a free-space silicon-based photon counter from Micro Photon Devices, with a peak efficiency of 40\% around 632 nm \cite{giudice2007}. The overall efficiency of the up-conversion unit, including the filter stage and the single-photon detector efficiency and coupling, is approximately 5\% at the maximum pump power. {\red In our experiment, the working point of the up-conversion detector was set to 2\% efficiency, in order to decrease the dark count rate generated by the intense pump laser}.  

\begin{figure}[ht!]
\centering
\includegraphics[width=0.8\textwidth]{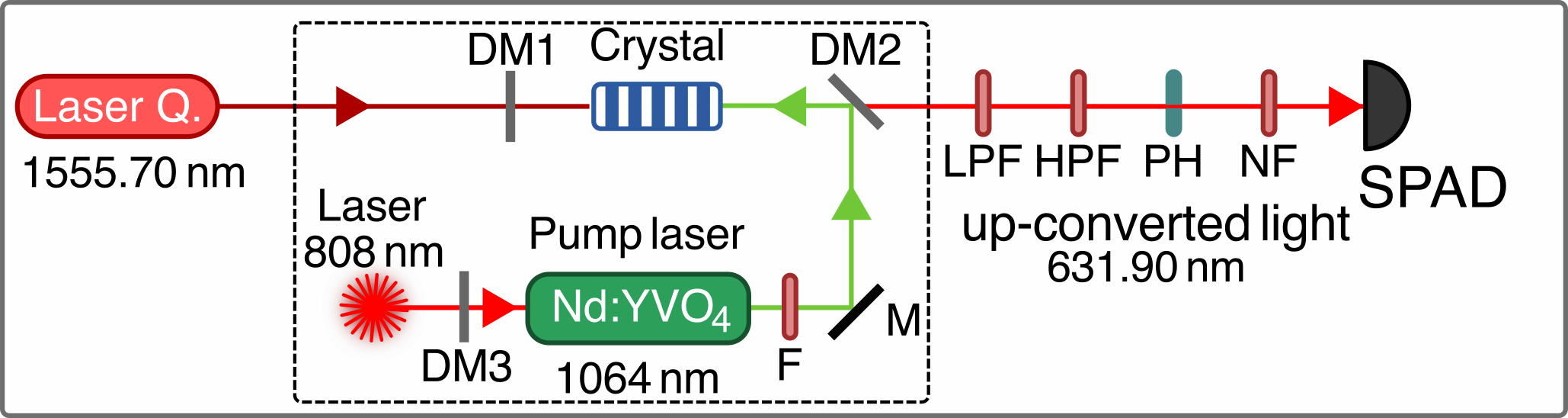}
\caption{\textbf{Detailed setup of the up-conversion scheme.} Laser Q: quantum laser, DM1,2,3: dichroic mirrors, M: mirror, LPF: low-pass filter, HPF: high-pass filter, PH: pinhole, NF: notch filters.}
\label{fig:up_conversion_scheme}
\end{figure}


\section*{Pulse width}
One of the key point of the single photon detector is the timing jitter. In the case of time-bin encoding, detection jitter is important to determine the maximum repetition rate of the source and to decrease the amount of timing errors at the receiver side. In this specific configuration, the properties of the up-conversion module combined with the free-space single photon detector allows for a very sharp pulse-shape, as reported in Figure \ref{fig:pulse_w}. In Figure \ref{fig:pulse_w} a) we reported the normalized count of our weak coherent state acquired by the InGaAs detector in non saturation regime and in the back-to-back configuration. The full-width-half maximum value is around 250 ps. On the contrary, the pulse acquired with the up-conversion unit combined with the silicon detector presents a FWHM value of about 130 ps, as reported in Figure \ref{fig:pulse_w} b). Thus, the silicon single photon detector allows an ultra precise time filtering technique, which limits the amount of spurious clicks in the time-bin window both from the dark counts event but also from the classical signals.

\begin{figure}[ht!]
\centering
\includegraphics[width=0.5\textwidth]{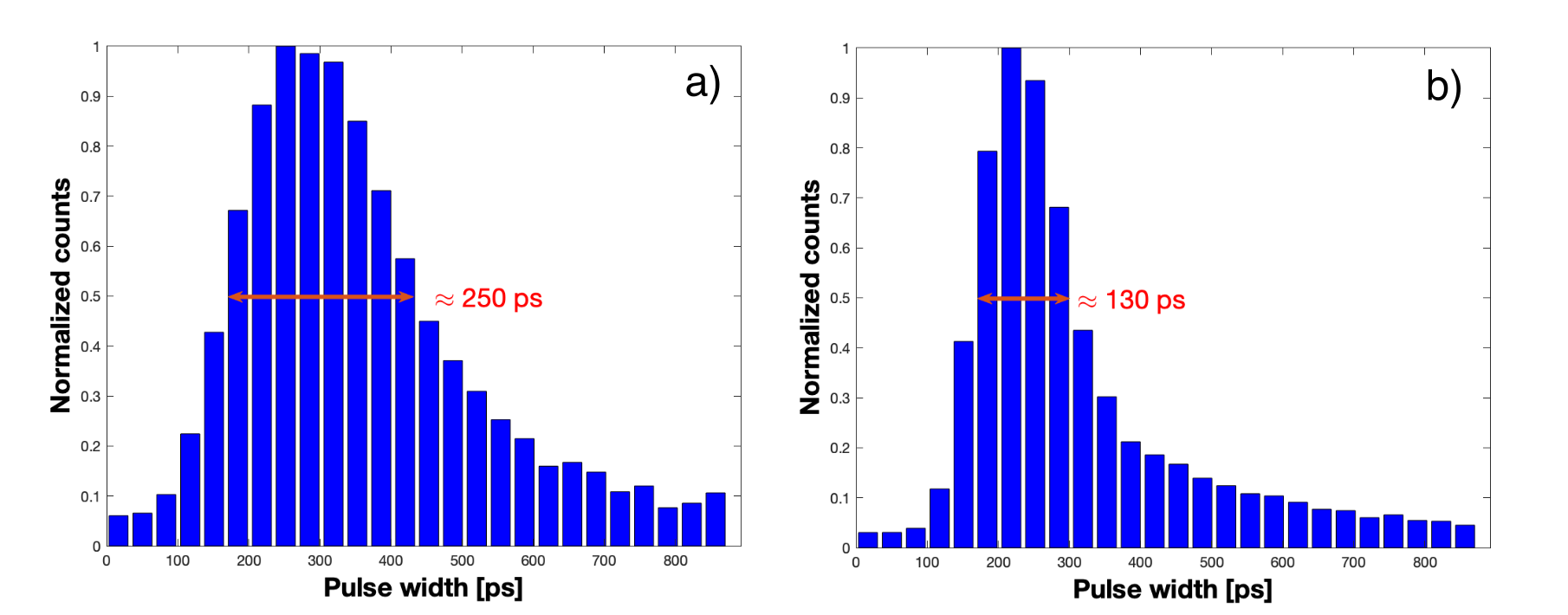}
\caption{\textbf{Pulse width acquired with two different single-photon detector}.}
\label{fig:pulse_w}
\end{figure}

\begin{table*}[t]
\setlength{\tabcolsep}{14pt}
\caption{{\bf Experimental parameters of the QKD test.} Here are reported the values that we set at the transmitter for each fiber channel, such as the mean photon number for signal and decoy states ($\mu_1$, $\mu_2$) and their relative probability ($p_{\mu_1}$). The state preparation rate is \SI{595}{\mega\hertz}, while the probability to prepare and measure the $\mathcal{Z}$~basis at the transmitter and the receiver are fixed to 90\% and 50\%, respectively, for both detection systems.}
\begin{center}
\begin{tabular}{|c|c|ccc|}
    \hline\hline
    &&&& \\[-0.4em] 
    \multirowcell{2}{{\bf quantum}\\{\bf channel}}
    & length & \SI{15}{\kilo\meter} & \SI{25}{\kilo\meter} & \SI{40}{\kilo\meter} \\[7pt]
    & loss & \SI{3}{\decibel} & \SI{5}{\decibel} & \SI{8}{\decibel} \\[7pt]
    \hline
    &&&& \\[-0.6em] 
    \multirowcell{4}{{\bf InGaAs} \\{\bf receiver} \\  ( 20\% efficiency,  \\ \SI{700}{\hertz} dark counts,\\
    \SI{20}{\micro\second} dead time)}
    & $\mu_1$ & 0.12 & 0.20 & 0.18 \\[8pt]
    & $\mu_2$ & 0.011 & 0.017 & 0.026  \\[8pt]
    & $p_{\mu_1}$ & 3\% & 3\% & 7\%  \\[8pt]     
    \hline
    &&&& \\[-0.6em] 
    \multirowcell{4}{{\bf Up-conversion} \\ {\bf receiver} \\  ( 2{\red{\sout{0}}}\% efficiency,  \\ \SI{11}{\kilo\hertz} dark counts, \\ \SI{77}{\nano\second} dead time)}
    & $\mu_1$ & 0.21 & 0.21 & / \\[8pt]
    & $\mu_2$ & 0.068 & 0.070 & / \\[8pt]
    & $p_{\mu_1}$ & 14\% & 19\% & / \\[8pt]    
    \hline\hline
\end{tabular}
\end{center}
\label{tab:1}
\end{table*}

\cleardoublepage

\end{document}